\documentclass{amsart}
\usepackage{amssymb,latexsym}
\usepackage{graphicx}

\begin{document}

\title{The Lamb-Bateman integral equation and the fractional derivatives}
\author{D. Babusci$^\dag$, G. Dattoli$^\ddag$, D. Sacchetti$^\diamond$} 

\address{$^\dag$ INFN - Laboratori Nazionali di Frascati, via E. Fermi 40, I-00044 Frascati.}
\email{danilo.babusci@lnf.infn.it}

\address{$^\ddag$ ENEA - Dipartimento Tecnologie Fisiche e Nuovi Materiali, Centro Ricerche Frascati\\
                 C. P. 65, I-00044 Frascati.}
 \email{giuseppe.dattoli@enea.it}
                 
\address{$^\diamond$ ENEA - Dipartimento di Statistica, Probabilit\`a e Statistica Applicata, Universit\`a 
                  "Sapienza" di Roma, P.le A. Moro, 5, 00185 Roma.}
\email{dario.sacchetti@uniroma1.it}

\begin{abstract}
The Lamb-Bateman integral equation was introduced to study the solitary wave diffraction and its solution 
was written in terms of an integral transform. We prove that it is essentially the Abel integral equation and 
its solution can be obtained using the formalism of fractional calculus.
\end{abstract}

\maketitle

This letter is motivated by an interesting article \cite{Mart} describing the life and achievements of Harry Bateman. 
The article reports, among the other things, the following integral equation proposed by Lamb \cite{Lamb} 
in his analysis of the diffraction of a solitary wave
\begin{equation}
\label{eq:lamb}
\int_0^\infty\,\mathrm{d}y\,u (x - y^2) \,=\, f (x)\;,
\end{equation}
where $ u (x)$ is a function to be determined. Either refs. \cite{Mart,Lamb} reports the solution suggested by Bateman 
without any proof. Here we show that relevant solution can be obtained using an operational formalism involving the 
fractional derivatives. 

By using the identity
\begin{equation}
\mathrm{e}^{\lambda \partial_x}\,g(x) \,=\, g (x + \lambda)\;
\end{equation}
eq. \eqref{eq:lamb} can be re-written as follows
\begin{equation}
\hat{O}\,u (x) \,=\, f (x)\;,
\end{equation}
where $\hat{O}$ is the following operator 
\begin{equation}
\hat{O} \,=\, \int_0^\infty\,\mathrm{d}y\,\mathrm{e}^{- y^2 \partial_x}\;.
\end{equation}
The use of the Gaussian integral 
\begin{equation}
\int_0^\infty\,\mathrm{d}z\,\mathrm{e}^{- a z^2} \,=\, \frac{1}{2}\,\sqrt{\frac{\pi}{a}}
\end{equation}
allows to write
\begin{equation}
\hat{O} \,=\, \frac{1}{2}\,\sqrt{\frac{\pi}{\partial_x}}
\end{equation}
provided that we assume the validity of the Gaussian integral for $a$ replaced by an operator, and, therefore, the solution 
of eq. \eqref{eq:lamb} can be written as the derivative of order 1/2 of the function $f (x)$, namely
\begin{equation}\label{eq:funux}
u (x) \,=\, \frac{2}{\sqrt{\pi}}\,\partial^{1/2} f (x)\;.
\end{equation}

The theory of fractional calculus \cite{OldSpa} allows the evaluation of  the so called \emph{differintegral} by means of an 
integral transform, namely
\begin{equation}
\partial^{- \mu} g(x) \,=\, \frac{1}{\Gamma (\mu)}\,\int_{- \infty}^x \,\mathrm{d}\xi\, g(\xi)\,(x - \xi)^{\mu - 1}\;,
\end{equation}
which in the case $\mu = 1/2$, yields
\begin{equation}
\partial^{- 1/2} g(x) \,=\, \frac{1}{\sqrt{\pi}}\,\int_{- \infty}^x \,\mathrm{d}\xi\, \frac{g(\xi)}{\sqrt{x - \xi}}\;.
\end{equation}
Accordingly, we recast eq. \eqref{eq:funux} in the form
\begin{equation}
u (x) \,=\, \frac{2}{\sqrt{\pi}}\,\partial^{- 1/2} \left[\partial_x f (x)\right]\;,
\end{equation}
and, thus\footnote{The change of variable is not necessary and was performed only  to get the same expression given by 
Bateman. As for the notation, we put $f^\prime (-\xi) = \partial_y f(y)|_{y = - \xi}$.} 
\begin{equation}
u (x) \,=\, \frac{2}{\pi}\,\int_{- x}^\infty \,\mathrm{d}\xi\, \frac{f^\prime (-\xi)}{\sqrt{x + \xi}}\;.
\end{equation}
which is the solution proposed by Bateman (see refs. \cite{Mart,Lamb} for further comments). 

Since this solution has been reported without any comment, it is not clear what procedure Bateman followed. Lamb in his paper 
reported a different solution in the form of a double integral, but added the remark: \emph{ Mr. H. Bateman, to whom I submitted the 
question, has obtained a simpler solution}. Nowadays the operational solution and the use of fractional derivatives looks natural,  
but it is not clear if these methods were used at the time Lamb wrote his paper. 

The procedure here outlined can also be exploited for integral equations that generalizes eq. \eqref{eq:lamb}, as,  for example
\begin{equation}
\label{eq:lambm}
\int_0^\infty\,\mathrm{d}y\,u (x - y^m) \,=\, f (x)\;.
\end{equation}
We write, indeed 
\begin{equation}
\hat{O}_m\,u (x) \,=\, f (x)\;,
\end{equation}
where
\begin{equation}
\hat{O}_m \,=\, \int_0^\infty\,\mathrm{d}y\,\mathrm{e}^{- y^m \partial_x} \,=\, 
\frac{1}{m}\,\Gamma \left(\frac{1}{m}\right)\,\partial_x^{- 1/m}\;,
\end{equation}
and the solution of eq. \eqref{eq:lambm} is given by
\begin{equation}
u (x) \,=\, \frac{m}{\Gamma \left(\frac{1}{m}\right)}\,\partial_x^{1/m} f(x)\;,
\end{equation}
which, finally, yields\footnote{We used the Euler reflection formula: $\Gamma (x) \Gamma (1 - x) = \frac{\pi}{\sin (\pi x)}$.}
\begin{equation}
u (x) \,=\, \mathrm{sinc}\left(\frac{\pi}{m}\right)\,\int_{- x}^\infty\,\mathrm{d}\xi\,\frac{f^\prime (-\xi)}{(x + \xi)^{1/m}}\;.
\end{equation}

Furthermore, the method can be extended to the study of the integral equation 
\begin{equation}
\int_{- \infty}^\infty\, \mathrm{d}y \, u( x - a y^2 + b y) \,=\, f(x)\;.
\end{equation}
In this case the operator $\hat{O}$ is
\begin{equation}
\hat{O} \,=\, \int_0^\infty\,\mathrm{d}y\,\mathrm{e}^{- y^2 \hat{A} + y \hat{B}} \,=\, 
\sqrt{\frac{\pi}{\hat{A}}}\, \exp \left(\frac{\hat{B}^2}{4 \hat{A}}\right)\;
\end{equation}
where $\hat{A} = a\,\partial_x$ and $\hat{B} = b\,\partial_x$. Therefore, we obtain
\begin{equation}
u (x) \,=\, \frac{\sqrt{a}}{\pi}\,\int_{- \infty}^x\,\mathrm{d}\xi\,\frac{f^\prime (\xi - c)}{\sqrt{x - \xi}}\;, 
\qquad\qquad \left(c \,=\, \frac{b^2}{4 a}\right)\;.
\end{equation}

We close this letter noticing that if we set $x - y^2 = z$ in eq. \eqref{eq:lamb}, we obtain
\begin{equation}
\int_0^\infty\,\mathrm{d}y\,u (x - y^2) \,=\, \frac{1}{2}\,\int_{- \infty}^x\,\mathrm{d}z\,\frac{u (z)}{\sqrt{x - z}} \,=\, f (x)\;, 
\end{equation}
that is the equation introduced by Abel in his study of tautochrone problem \cite{Abel}, and that paved the way to the theory 
of integral equations.

\vspace{0.5cm}
\section*{Acknowledgements}
The authors wish to thank prof. Kazuyuki Fujii for having pointed out some misprints.

\end{document}